\documentclass[fleqn,usenatbib]{mnras}

\usepackage{newtxtext,newtxmath}

\usepackage[T1]{fontenc}

\DeclareRobustCommand{\VAN}[3]{#2}
\let\VANthebibliography\thebibliography
\def\thebibliography{\DeclareRobustCommand{\VAN}[3]{##3}\VANthebibliography}

\usepackage{graphicx}	
\usepackage{amsmath}	

\title[V618 Sgr: Galactic eclipsing symbiotic nova]{V618 Sgr: Galactic eclipsing symbiotic nova detected in repeated outbursts}

\author[J. Merc et al.]{
J.~Merc,$^{1}$\thanks{E-mail: jaroslav.merc@mff.cuni.cz}
R.~G\'{a}lis,$^{2}$
P.~Velez,$^{3}$
S.~Charbonnel,$^{4}$
O.~Garde,$^{4}$
P.~Le~D\^u,$^{4}$        
L.~Mulato,$^{4}$
T.~Petit,$^{4}$
\newauthor
T.~Bohlsen,$^{3}$
S.~Curry,$^{3}$
T.~Love,$^{3}$
H.~Barker$^{3}$
\\
$^{1}$Astronomical Institute, Faculty of Mathematics and Physics, Charles University, V Hole\v{s}ovi\v{c}k{\'a}ch 2, 180 00 Prague, Czechia\\
$^{2}$Institute of Physics, Faculty of Science, P. J. \v{S}af{\'a}rik University, Park Angelinum 9, 040 01 Ko\v{s}ice, Slovakia\\
$^{3}$Astronomical Ring for Amateur Spectroscopy Group (ARAS)\\
$^{4}$Southern Spectroscopic Project Observatory Team (2SPOT), 45 Chemin du Lac, 38690 Ch\^{a}bons, France
}

\date{Accepted 2023 May 03. Received 2023 April 30; in original form 2023 February 24}

\pubyear{2023}

\begin{document}
\label{firstpage}
\pagerange{\pageref{firstpage}--\pageref{lastpage}}
\maketitle

\begin{abstract}
V618 Sgr was previously classified as an R CrB-type variable and later as a possible symbiotic star. Our study aims to analyse the nature of this target, which is currently undergoing significant brightening in properties similar to those of known symbiotic novae. We analyse literature information, photometric observations, and 35 new optical spectra. Our findings strongly suggest that V618 Sgr is an eclipsing symbiotic nova currently in outburst. Additionally, since the star has demonstrated at least two similar brightenings in the past, we propose that V618 Sgr could be the first known galactic symbiotic nova observed in repeated outbursts of this type and may host a relatively massive white dwarf.
\end{abstract}




\section{Introduction}
V618 Sgr (= 2MASS J18075721-3629523; $\alpha_{\rm 2000}$ = 18:07:57.21, $\delta_{\rm 2000}$ = -36:29:52.30) was initially classified as an R~CrB variable on the basis of its photometric behaviour \citep[][]{1940AnHar..90..207S}. This classification was later contradicted by \citet{1989Obs...109..229K}, since his spectrum covering the spectral region of 3\,500\,-\,5\,500\,\AA\,\,showed prominent TiO bands and several emission lines (especially the Balmer lines of hydrogen and singly ionised iron). Since R~CrB-type stars are hydrogen deficient, \citet{1989Obs...109..229K} suggested that V618 Sgr belongs to the symbiotic group. On the other hand, emission lines with a higher ionisation potential (e.g., [O\,III], He\,II, [Fe\,VII]), typical for symbiotic binaries, were missing in the observed spectrum, and singly ionised helium lines were also absent (or at least very weak), preventing definite confirmation of the symbiotic nature of V618~Sgr. Therefore, it was only included among the symbiotic candidates in the catalogues of \citet{2000A&AS..146..407B} and \citet{2019ApJS..240...21A} and has not received any significant attention from the symbiotic community yet.

We have included V618 Sgr to our New Online Database of Symbiotic Variables\footnote{\url{http://astronomy.science.upjs.sk/symbiotics/}} \citep{2019RNAAS...3...28M} and selected it, among other objects, for the observing campaign focused on the poorly studied symbiotic stars and candidates \citep[see also][]{2020MNRAS.499.2116M,2021MNRAS.506.4151M,2022MNRAS.510.1404M}. In this paper, we present the results of the analysis of the photometry collected from various sources and our new spectroscopic observations, supplemented by the available information on this target from the literature. Based on our results, we conclude that V618 Sgr is indeed a symbiotic binary that is currently undergoing a symbiotic nova outburst. Moreover, given that similar brightenings have been observed in the past, it is very probable that V618 Sgr can be classified as the first known galactic symbiotic nova observed in repeated outbursts of this type.

\begin{figure*}
   \centering
   \includegraphics[width=1.88\columnwidth]{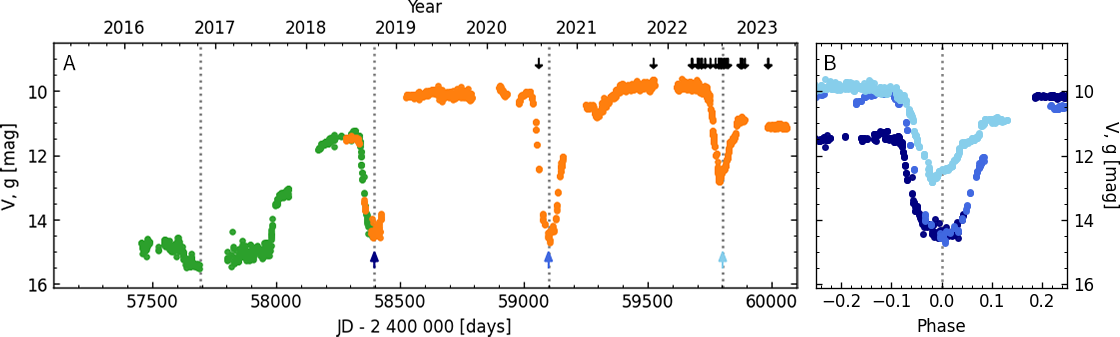}
      \caption{\textbf{A:} Recent light curve of V618 Sgr constructed on the basis of ASAS-SN data ($V$ and $g$ observations are shown in green and orange, respectively). The light curve has been processed in order to remove the apparent two brightness levels (see the text for more details). The dotted lines denote the times of eclipses calculated using Eq. \ref{eq:eph}. Black arrows at the top denote the times when our spectra were obtained, and blue-colored arrows depict by which color the particular eclipse is shown on the adjacent panel. \textbf{B:} Part of the phased light curve ($P = 703$\,d) showing the eclipses.}
         \label{fig:LC}
\end{figure*}

\begin{figure}
   \centering
   \includegraphics[width=0.95\columnwidth]{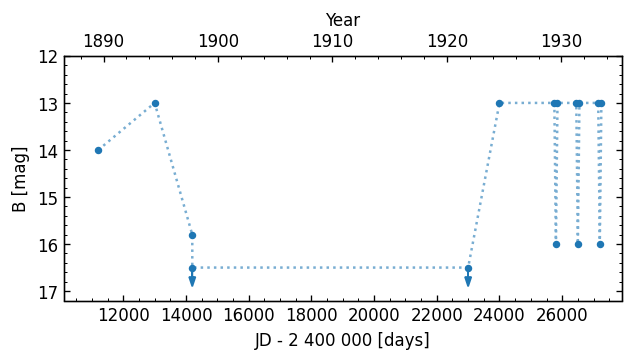}
      \caption{Light curve of V618 Sgr between 1889 and 1933 reconstructed based on the textual description provided by \citet[][]{1940AnHar..90..207S}. The arrows denote the upper limits.}
         \label{fig:historical}
\end{figure}

\section{Observations}
To study the photometric evolution of V618 Sgr, we collected data obtained by the All-Sky Automated Survey for Supernovae \citep[ASAS-SN; \textit{V}~and \textit{g} filters; ][]{2014ApJ...788...48S, 2017PASP..129j4502K} covering JD\,2\,457\,456 -- 2\,460\,060 (March 9, 2016 -- April 25, 2023) and supplemented them with observations submitted by amateur observers into the database of the American Association of Variable Star Observers\footnote{\url{https://www.aavso.org/}} \citep[AAVSO; visual observations and data in \textit{V} filter; ][]{kafka}. These data cover the interval from JD\,2\,443\,721 to JD\,2\,459\,776 (July~31, 1978 -- July 15, 2022). In order to analyse the behaviour of V618 Sgr in the past, we searched for the photometry published in the literature. Although no light curve of V618 Sgr has ever been published, we analysed the description of the behaviour presented by \citet{1940AnHar..90..207S} and \citet{1975PVSS....3....1B}.

The spectroscopic observations of V618 Sgr presented in this study were obtained primarily by small telescopes in the scope of our observing campaign coordinated with the ARAS Group\footnote{\url{https://aras-database.github.io/database/symbiotics.html}} \citep[\textit{Astronomical Ring for Amateur Spectroscopy};][]{2019CoSka..49..217T}. Between JD\,2\,459\,061 and JD\,2\,459\,992 (July 30, 2020 -- February 16, 2023), we collected 35 low-resolution spectra of V618 Sgr ($R \sim 500 - 3\,000$) covering various regions of the optical part of the spectrum. These data were obtained using equipment in Chile (30-cm Ritchey-Chretien telescope, Alpy600 spectrograph), New Zealand (20-cm Newtonian telescope with L200 Littrow spectrograph; 30-cm Ritchey-Chretien telescope with Alpy600 spectrograph), and Australia (32-cm Planewave CDK telescope, UVEX spectrograph; 28-cm Celestron telescope, LISA spectrograph). The log of observations is available in Tab. \ref{table:log_obs}.

To construct the multi-frequency spectral energy distribution (SED) of V618 Sgr, we have collected data from the \textit{Gaia} DR3 \citep[\textit{$G_{\rm BP}$, \textit{G}, \textit{$G_{\rm RP}$}};][]{2022arXiv220800211G}, the AAVSO Photometric All-Sky Survey \citep[APASS; \textit{B, V;}][]{2015AAS...22533616H}, the Two Micron All-Sky Survey \citep[2MASS; \textit{J, H, K};][]{2006AJ....131.1163S}, and \textit{the Wide-field Infrared Survey Explorer} \citep[\textit{WISE}; \textit{W1, W2, W3, W4};][]{2010AJ....140.1868W}.

\section{Photometric behaviour}
\subsection{Recent light curve}
The recent light curve of V618 Sgr, constructed on the basis of observations from the ASAS-SN survey, is depicted in Fig.~\ref{fig:LC}A. The apparent outliers were removed from the light curve. In addition, the light curve, as processed by the ASAS-SN Sky Patrol server, showed two apparent brightness levels, particularly visible in the period of 2021 -- 2022. We have corrected for this seemingly processing effect by shifting one of the levels to the brightness of another. The shift was found by minimising the $\chi^2$ value. The correct brightness has been verified with the help of the AAVSO observations. 

In general, the target light curve shows a prominent brightening with the onset in 2017 (around JD\,2\,457\,975) that went unnoticed by observers and was not reported in the literature. In a way, this is to be expected as V618 Sgr is a known variable star\footnote{Classified as R CrB-type in the SIMBAD database \citep{2000A&AS..143....9W} and the General Catalogue of Variable Stars \citep{2017ARep...61...80S}, even though this classification was rejected a long time ago; see the Introduction.} and most surveys do not report on the brightness changes of stars that are expected to vary, and only publish new transients.

It is apparent from the light curve that the rise to maximum brightness took about 1.5~years, and the brightening had an approximate amplitude of 5\,mag in the $V$ filter. The star has been at its maximum brightness since 2019 at least until middle 2022. Very recent data suggest that we might already be witnessing a gradual decrease in brightness after the plateau phase. 

On three occasions, significant drops in brightness have been recorded (Fig.~\ref{fig:LC}A). These drops repeated with a period of 703 days, and we attributed them to eclipsing events. The linear ephemeris for the mid-eclipse would be:

\begin{equation}
\label{eq:eph}
JD{\rm_{min}} = 2\,458\,397\pm8 + 703\pm4 \times E  
\end{equation}

The recent light curve phased with the obtained orbital period is shown in Fig. \ref{fig:LC}B. The significant difference between the shapes of the 2018 and 2020 eclipses and the one observed in 2022 is obvious. This change corresponds to the notable transition in the spectroscopic appearance of V618 Sgr (see the discussion in Sec. \ref{sec:spec}).

The recent photometric behaviour resembles the brightness changes observed in 'slow' symbiotic novae \citep[not to be confused with so-called 'symbiotic recurrent novae'; see more in][]{2010arXiv1011.5657M,2019arXiv190901389M}. The light curves of these systems are characterised by a slow rise to maximum brightness (amplitude of 3 -- 7 magnitudes within a few years), followed by a prolonged decline during several decades \citep[the duration of the outburst is inversely proportional to the mass of the white dwarf; see fig. 6 in][]{2010arXiv1011.5657M}. 

The group of symbiotic stars that manifest this type of thermonuclear nova outbursts is relatively small (contains only about a dozen systems\footnote{See the up-to-date list of symbiotic stars detected in outbursts in the New Online Database of Symbiotic Variables: \url{http://astronomy.science.upjs.sk/symbiotics/utilities/outbursts.html}.}; well-known examples include AG Peg, RR Tel, HM Sge, V1016 Cyg, or PU Vul). Moreover, we attributed the drops in brightness observed in V618 Sgr to the eclipses of the hot component in the outburst, which would make this system a member of an even more exclusive group of eclipsing symbiotic novae. 

The detected eclipses (those in 2018 and 2020) can be used to estimate the lower limits of radii of the binary components around the optical maximum. If we assume that the orbit of V618 Sgr is circular, $i$ = 90$^{\circ}$, and the total mass of the system is $\sim$\,2 - 3\,M$_\odot$ \citep[i.e., similar to those of known symbiotic stars; see, e.g.,][]{2003ASPC..303....9M}, the contact times (t$_4$ - t$_1$ $\sim$ 125\,d; \mbox{t$_3$ - t$_2$ $\sim$ 27\,d)} gives us the sizes of the eclipsing giant and the eclipsed hot component of $R_{\rm G}$ = 142 -- 163 R$_\odot$ and $R_{\rm h}$ = 92 -- 105 R$_\odot$, respectively. The derived radius of the eclipsing giant is very well consistent with an M6-7 III star \citep{1999AJ....117..521V}, matching the results of the analysis of our low-resolution spectra and SED (see Sec. \ref{sec:spec}). The size of the eclipsed object is very similar to the hot components of other symbiotic novae during the outbursts \citep[e.g., PU Vul, see][]{2012ApJ...750....5K}. 

The third of the observed eclipses (2022) show significant changes in the sizes (probably even appearance of density structures). Assuming that the radius of the cool giant has not changed significantly, the observed changes would be consistent with the shrinkage of the hot component, as often observed in symbiotic novae in outburst \citep[][]{2012ApJ...750....5K}.

\subsection{Historical evolution}

It is very interesting to compare the recent light curve with the photometric evolution observed by \citet[][]{1940AnHar..90..207S} and interpreted as R~CrB behaviour. Since no light curve was published in the paper (although the existence of 143 measurements before 1924 is mentioned), we reconstructed it based on the limited provided textual description (Fig.~\ref{fig:historical}). V618 Sgr showed two brightenings in the period of 1889 -- 1933. During the latter period of increased brightness, three minima were observed, which were repeated at an interval of $\sim$ 700\,days. This is the same period that we inferred for the eclipses observed in the recent light curve (703\,days). 

In general, at least during the second brightening period, the photometric behaviour resembles the currently observed outburst of V618~Sgr (the brightness remaining at approximately the same maximum level over the years with occasional interruptions by the eclipses). It is probable that this brightening was also a nova outburst of V618 Sgr. The fact that \citet{1989Obs...109..229K} did not detect emission lines with high ionisation potential in his spectrum obtained in October 1988, more than 60 years after this outburst, would suggest that the giant in V618 Sgr was not capable of sustaining stable shell burning in the long term. The white dwarf entered an accreting-only state and started accreting matter in the nova regime again \citep[see more details on such an evolution in][]{2020A&A...636A..77S}. That gave rise to the currently observed symbiotic nova outburst.

The only other information on the photometric evolution of V618~Sgr available in the literature is the non-detection of the system during 242 visual observations of A. F. Jones obtained between 1956 and 1972 and reported by \citet{1975PVSS....3....1B}. The upper limit mentioned for these observations was 13.5 mag which is fainter than the maximal brightness achieved during the recent and apparently also during both historical maxima. From this we conclude that during the observed time interval, there was probably no outburst of V618 Sgr. More or less regular visual observations of V618 Sgr by AAVSO observers started in 1978, and until the recent brightening, no significant variability was seen in the light curve. Part of this time interval is also covered by the observations of the All-Sky Automated Survey \citep[ASAS;][]{1997AcA....47..467P} that confirm that the target was in the low state. If there were indeed no outbursts between the last one observed by \citet[][]{1940AnHar..90..207S} and the recent one, this would suggest a recurrence time of about 90 years. On the other hand, the quiescence between the two brightenings observed by \citet[][]{1940AnHar..90..207S} was significantly shorter. In addition, the decrease in brightness after the first one appears to be relatively rapid, but without spectroscopic information, the reason for that remains unclear.

\section{Spectroscopic appearance and the SED}\label{sec:spec}
\subsection{Cool component and the distance}

\begin{figure}
   \centering
   \includegraphics[width=\columnwidth]{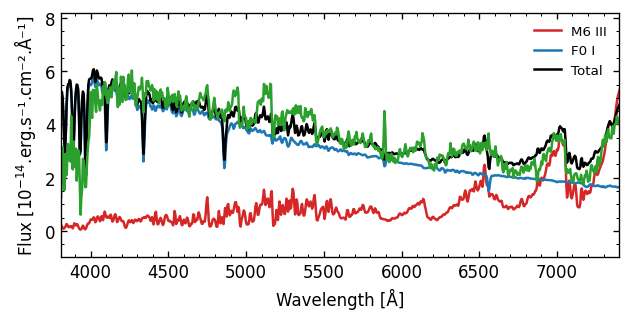}\vspace{-4mm}
      \caption{Optical dereddened spectrum of V618 Sgr (shown in green) obtained during the ingress of the eclipse in 2020 (JD 2\,459\,061) clearly showing the contribution of the cool star. The spectrum was calibrated to the approximate flux in $g$ filter using the nearly-simultaneous ASAS-SN observations. The cool giant and the outbursting component are approximated by the spectra from the library of \citet{1998PASP..110..863P} and are shown in red and blue, respectively.}
         \label{fig:eclipse_spec}
\end{figure}

We have acquired 35 optical spectra of V618 Sgr. The times when the spectra were obtained are indicated by arrows in Fig. \ref{fig:LC}A. By chance, our first spectrum was obtained during the eclipse of the system in 2020. In the spectrum (Fig.~\ref{fig:eclipse_spec}), the presence of the cool star is well visible thanks to the prominent TiO bands. This is consistent with the spectroscopic observations of \citet{1989Obs...109..229K} obtained during the quiescence in 1988 showing strong TiO bands.

\begin{figure}
   \centering
   \includegraphics[width=\columnwidth]{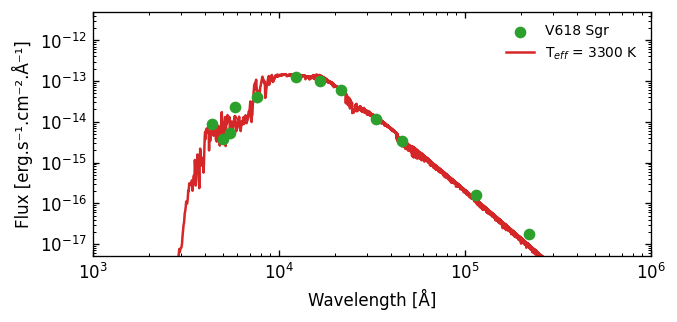}\vspace{-4mm}
      \caption{Multi-frequency SED of V618 Sgr constructed on the basis of data from \textit{Gaia} DR3, APASS, 2MASS, and \textit{WISE} (green symbols). The fluxes were de-reddened by $E_{\rm (B-V)}$ = 0.14 mag using the reddening law of \citet{1989ApJ...345..245C} and adopting the total to selective absorption ratio $R = 3.1$. The best fitting theoretical spectrum \citep{2014IAUS..299..271A} is plotted in red.}
         \label{fig:sed}
\end{figure}

The late-type spectral classification is further confirmed by the analysis of the SED of V618 Sgr (Fig.~\ref{fig:sed}) constructed on the basis of pre-outburst observations. It is consistent with a cool star with temperature $T_{\rm eff}$ $\sim$ 3\,300\,K \citep[$\sim$\,M6 giant;][]{1999AJ....117..521V}. We should note that the data used in our SED fitting were not observed simultaneously, so one should be cautious when interpreting the results. Although there seems to be some variability present in the system outside of the outbursts (seen in the ASAS-SN and \textit{Gaia} data; with an amplitude of a few tenths of magnitude), it would not significantly influence the results. We tested fitting the SED of V618 Sgr constructed on the basis of various combinations of data from several surveys using the VO SED analyser (VOSA) on the Spanish Virtual Observatory theoretical services website \citep{2008A&A...492..277B}\footnote{http://svo2.cab.inta-csic.es/theory/vosa/} and all of this resulted in similar values of effective temperature in the range of 3\,100 -- 3\,300\,K. A~similar temperature (3\,255\,K) was also obtained by \citet{2022A&A...658A..91A} using the StarHorse code. In addition, it is worth noting at this point that the SED does not show the presence of the hot component. However, given that the system was most likely in the accreting-only state (as discussed in the previous section), the hot component would be too faint at these wavelengths to contribute significantly.

Adopting the distance of 5.2 kpc (with an uncertainty of 4.5 -- 6.2\,kpc) obtained by \citet{2021AJ....161..147B} using the Gaia EDR3 data, $G$ magnitude ($G$ = 13.0 mag) and $G_{\rm BP}$-$G_{\rm RP}$ colour ($G_{\rm BP}$-$G_{\rm RP}$ = 4.1 mag) from Gaia DR3 \citep{2022arXiv220800211G}, and the extinction $E_{\rm (B-V)}$ = 0.14 mag \citep{2011ApJ...737..103S} would safely position the cool component of V618 Sgr among giant stars in the Gaia HR diagram \citep[see fig. 5 in][]{2018A&A...616A..10G} and rule out its possible main-sequence classification - the absolute magnitudes of the giant and the main-sequence star would differ by about 15 magnitudes for such a cool star. 
The StarHorse catalogue \citep{2022A&A...658A..91A} gives a slightly higher distance of about 5.8 kpc (with an uncertainty of 5.5 -- 6.2\,kpc) than the one reported by \citet{2021AJ....161..147B} and a higher extinction value of $E_{\rm (B-V)}$ = 0.29 mag (0.25 -- 0.34 mag). 

We also employed independent approaches to obtain the distance of V618~Sgr. \citet[][]{2005A&A...440..995S} presented an empirical relation between the observed bolometric flux and the distance. Using the relation for red giants \citep[eq. 27 in][]{2005A&A...440..995S} and the bolometric flux from our SED fit (see below), we obtained a distance of about 4.9 kpc. Another estimate of the distance comes from the comparison of the observed and reddened $K$ magnitude of V618~Sgr from the 2MASS catalogue ($K$ = 7.10 -- 7.15 mag for the reddening values above) and the absolute $K$ magnitude of the mid-M giant (M4-6; $M_{\rm K}$ = -6.25 -- -5.55 mag) calculated using the luminosities from \citet{1987A&A...177..217D} and bolometric corrections from \citet{2010MNRAS.403.1592B}. This approach resulted in a distance of 3.4 -- 4.8\,kpc. This range is slightly lower than the ones obtained from astrometry.

For the rest of the paper, we have adopted the distance of 5.2\,kpc as presented by \citet{2021AJ....161..147B}. In any case, for any distance in the range mentioned above, the conclusion that the donor in V618 Sgr is a giant, not a main-sequence star holds. The main-sequence star with this apparent magnitude would need to be at a distance of $\sim$ 20 pc.

\begin{figure}
   \centering
   \includegraphics[width=\columnwidth]{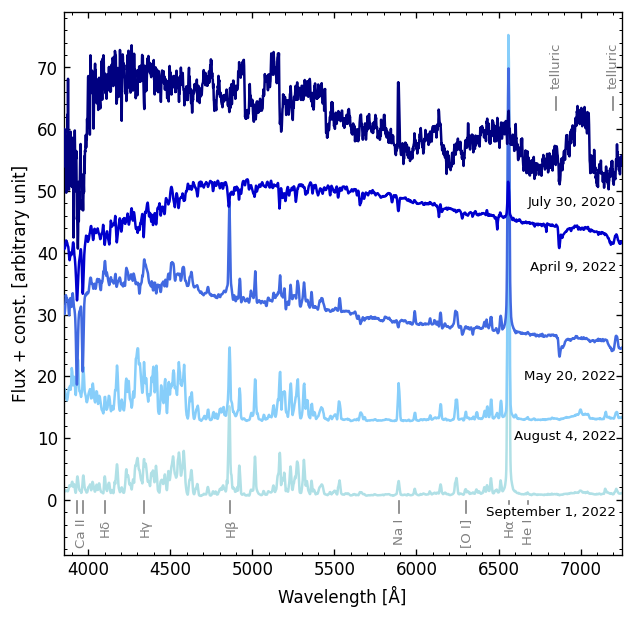}\vspace{-4mm}
      \caption{Evolution of the optical spectra of V618 Sgr. Identification of prominent emission lines is given at the bottom of the figure. The majority of the unlabeled emission lines are of Fe\,{\sc II}.}
         \label{fig:spectra}
\end{figure}

\subsection{Spectroscopic evolution throughout the outburst}
\begin{figure}
   \centering
   \includegraphics[width=\columnwidth]{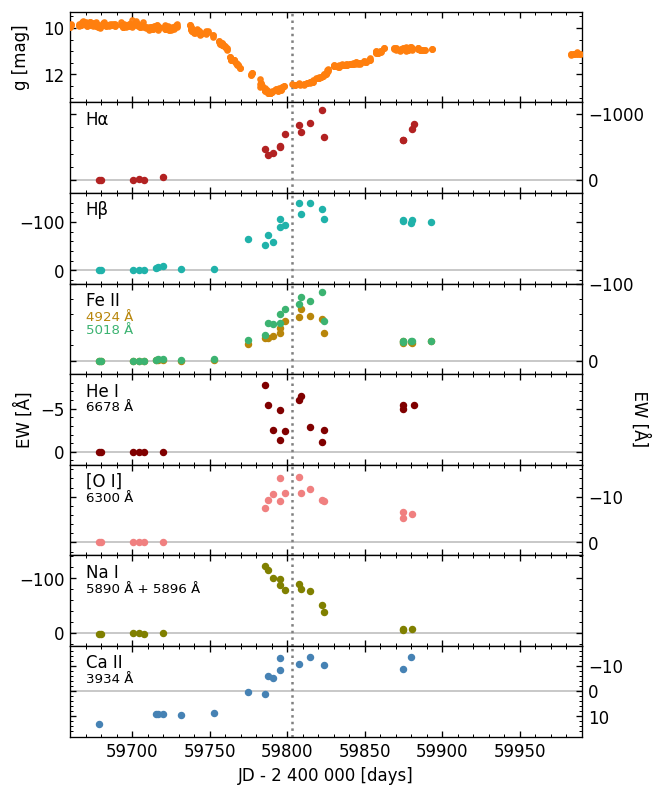}\vspace{-4mm}
      \caption{Evolution of equivalent widths of selected emission lines in the spectrum of V618 Sgr. The top panel shows the light curve in the $g$ filter for comparison. The dotted line denotes the time of mid-eclipse calculated using Eq. \ref{eq:eph}.}
         \label{fig:ews}
\end{figure}

While during the eclipse in 2020, the signatures of the cool star are visible in the optical spectra (Fig. \ref{fig:spectra}), after the eclipse until mid-May 2022, V618 Sgr had the spectroscopic appearance of a typical symbiotic nova after the optical maximum showing the continuum shape of an \mbox{F} type star \citep[typical for symbiotic novae in outbursts;][]{2019arXiv190901389M}. Only H$\alpha$ was seen in the emission (see the spectrum from April 2022 in Fig.~\ref{fig:spectra}). The continuation of our monitoring revealed the significant change in spectral appearance that occurred in May 2022 (sometime between May 8 and May 16). Several emission lines appeared in the spectrum (especially several prominent Balmer lines of H\,{\sc I} and many Fe\,{\sc II} lines; Fig.~\ref{fig:spectra}). In June 2022, the brightness of V618 Sgr started to decrease during the eclipse, and Na I and [O I] appeared in emission. Since the middle of July 2022, the Ti\,II and He\,I lines, and possibly also the N III, [Fe IV] lines, have been detectable in the spectrum. At around the same time, Ca II switched from absorption to strong emission. Especially in July and August 2022 (when V618 Sgr was in the middle of the eclipse), mild TiO bands appeared in the red region of the spectra. In the spectrum obtained after solar conjunction in February 2023, the only remarkable changes were the weakening of Na I lines and the appearance of [N\,II]. The evolution of the equivalent widths of the selected emission lines is shown in Fig. \ref{fig:ews}. Notable is the increase in the strength of the lines during the eclipse in 2022.

The observed evolution of the optical spectrum of V618 Sgr is common for symbiotic novae. A very similar abrupt appearance of the H I and Fe II emission lines, as detected by us in V618 Sgr, was also observed in the case of known symbiotic nova RR Tel \citep{1977MmRAS..83....1T}. The author also reported a slow increase in the maximal ionisation potential of the emission lines in the spectrum, and we may also see the beginning of that slow increase in our case. The evolution of other symbiotic novae, V4368 Sgr \citep[well documented in fig. 6.5 in][]{2019arXiv190901389M} and PU Vul \citep[fig. 1 in][]{2018MNRAS.479.2728C}, also seems to be very similar to what we have observed for V618 Sgr up to now.

The evolution of spectra and the appearance of emission lines also allow us to approximately trace the changes in the temperature of the hot component. The first spectrum we obtained during the eclipse in 2020 shows the presence of a shell spectrum similar to the $\sim$F0 I star \citep[temparature of about 7\,700\,K;][]{1998PASP..110..863P}. In April 2022, the spectrum seemed to be even cooler ($\sim$F7; 6\,300\,K). Later, the appearance of emission lines suggests an increase in the hot component temperature. The lower limit can be obtained from the maximum ionisation potential $IP$ derived from the spectrum and using the relation $T\rm_{h}$ [10$^3$\,K] $\sim$ $IP_{\text{max}}$\,[eV] \citep{1994A&A...282..586M}. This method suggests that the temperature increased to about 7\,900\,K when the Fe II lines appeared, and at least to $\sim$14\,500\,K at the time when the [N II] lines were detected, maybe even to $\sim$30\,000\,K if the detection of the N III and [Fe IV] lines is real. The changes in the temperature of the hot component are connected with the contraction of its radius \citep[see, e.g., fig 3 in][]{2012ApJ...750....5K}.

Finally, we can also get a rough estimate of the energy radiated in the outburst. Assuming that the luminosity of the outbursting component was constant throughout the plateau phase, we can convert the observed $V$ magnitudes to absolute bolometric magnitudes using the bolometric correction from \citet{1996ApJ...469..355F} and the adopted distance of 5.2\,kpc. From them, we can calculate the luminosity, and by integrating over time, we can estimate the radiated energy of about 0.5\,$\times10^{46}$erg. This is at the lower end of the range given by \citet{1994A&A...282..586M} for other symbiotic novae. However, this value is affected by inaccuracies in both distance and reddening, and at present we are only able to determine the lower limit of energy released as the outburst is not yet over.

\section{Conclusions}
The data analysed in this study support the symbiotic nova classification of V618~Sgr. This makes the object very interesting because it was already observed in at least two outbursts in the past. This would make V618 Sgr the first confirmed (and only, for now) galactic recurrent 'slow' symbiotic nova. The only two comparable objects might be a known symbiotic nova RX Pup, detected in the outburst in the 1970s with indecisive evidence about another outburst that occurred around 1894 \citep{1999MNRAS.305..190M}, and a symbiotic star LMC~S154 located in the Large Magellanic Cloud. \citet{2019A&A...624A.133I} suggested that this system has experienced at least one, but more likely two or three symbiotic nova outbursts in the past. However, the evidence of multiple outbursts is not as straightforward as in the case of V618 Sgr, and the outbursts of LMC S154 might have been more similar to those of symbiotic recurrent novae, not 'slow' symbiotic novae.

The analysed light curves of V618 Sgr confirmed that this system manifests eclipses. The orbital period (703 days) is typical for S-type symbiotic stars \citep[see, e.g.,][]{1999A&AS..137..473M,2012BaltA..21....5M,2013AcA....63..405G}. Such an infrared classification is also supported by the NIR infrared colours of V618 Sgr. We also note that V618 Sgr satisfies the symbiotic IR criteria and classification trees of \citet[][]{2019MNRAS.483.5077A,2021MNRAS.502.2513A}.

Finally, the relatively short recurrence time of outbursts of V618~Sgr suggests that the white dwarf in this system could be quite massive \citep[e.g.,][]{2007ApJ...663.1269N,2013ApJ...777..136W}. This is further supported by the relatively short duration of the plateau phase in the current outburst (probably < 4 years), as the duration of this stage is inversely proportional to the mass of the white dwarf \citep[fig. 6 in][]{2010arXiv1011.5657M}. Therefore, it would be interesting to follow the evolution of the system in the next years/decades, as it could be faster in comparison with typical 'slow' symbiotic novae. That makes V618 Sgr a good testbed for studying such nova outbursts.

\section*{Acknowledgements}
We are thankful to an anonymous referee for the comments and suggestions improving the manuscript. We acknowledge with thanks the variable star observations from the AAVSO International Database contributed by observers worldwide and used in this research. This research was supported by the \textit{Charles University}, project GA UK No. 890120, and the Slovak Research and Development Agency under contract No. APVV-20-0148.

\section*{Data Availability}
 
The spectra used in the paper are available in the ARAS Database. Photometric data are accessible from the websites of the surveys. The other data are available on reasonable request to the authors.



\bibliographystyle{mnras}
\bibliography{example} 

\begin{thebibliography}{}
\makeatletter
\relax
\def\mn@urlcharsother{\let\do\@makeother \do\$\do\&\do\#\do\^\do\_\do\%\do\~}
\def\mn@doi{\begingroup\mn@urlcharsother \@ifnextchar [ {\mn@doi@}
  {\mn@doi@[]}}
\def\mn@doi@[#1]#2{\def\@tempa{#1}\ifx\@tempa\@empty \href
  {http://dx.doi.org/#2} {doi:#2}\else \href {http://dx.doi.org/#2} {#1}\fi
  \endgroup}
\def\mn@eprint#1#2{\mn@eprint@#1:#2::\@nil}
\def\mn@eprint@arXiv#1{\href {http://arxiv.org/abs/#1} {{\tt arXiv:#1}}}
\def\mn@eprint@dblp#1{\href {http://dblp.uni-trier.de/rec/bibtex/#1.xml}
  {dblp:#1}}
\def\mn@eprint@#1:#2:#3:#4\@nil{\def\@tempa {#1}\def\@tempb {#2}\def\@tempc
  {#3}\ifx \@tempc \@empty \let \@tempc \@tempb \let \@tempb \@tempa \fi \ifx
  \@tempb \@empty \def\@tempb {arXiv}\fi \@ifundefined
  {mn@eprint@\@tempb}{\@tempb:\@tempc}{\expandafter \expandafter \csname
  mn@eprint@\@tempb\endcsname \expandafter{\@tempc}}}

\bibitem[\protect\citeauthoryear{{Akras}, {Guzman-Ramirez}, {Leal-Ferreira}  \&
  {Ramos-Larios}}{{Akras} et~al.}{2019a}]{2019ApJS..240...21A}
{Akras} S.,  {Guzman-Ramirez} L.,  {Leal-Ferreira} M.~L.,   {Ramos-Larios} G.,
  2019a, \mn@doi [\apjs] {10.3847/1538-4365/aaf88c}, \href
  {https://ui.adsabs.harvard.edu/abs/2019ApJS..240...21A} {240, 21}

\bibitem[\protect\citeauthoryear{{Akras}, {Leal-Ferreira}, {Guzman-Ramirez}  \&
  {Ramos-Larios}}{{Akras} et~al.}{2019b}]{2019MNRAS.483.5077A}
{Akras} S.,  {Leal-Ferreira} M.~L.,  {Guzman-Ramirez} L.,   {Ramos-Larios} G.,
  2019b, \mn@doi [\mnras] {10.1093/mnras/sty3359}, \href
  {https://ui.adsabs.harvard.edu/abs/2019MNRAS.483.5077A} {483, 5077}

\bibitem[\protect\citeauthoryear{{Akras}, {Gon{\c{c}}alves}, {Alvarez-Candal}
  \& {Pereira}}{{Akras} et~al.}{2021}]{2021MNRAS.502.2513A}
{Akras} S.,  {Gon{\c{c}}alves} D.~R.,  {Alvarez-Candal} A.,   {Pereira} C.~B.,
  2021, \mn@doi [\mnras] {10.1093/mnras/stab195}, \href
  {https://ui.adsabs.harvard.edu/abs/2021MNRAS.502.2513A} {502, 2513}

\bibitem[\protect\citeauthoryear{{Allard}}{{Allard}}{2014}]{2014IAUS..299..271A}
{Allard} F.,  2014, in {Booth} M.,  {Matthews} B.~C.,   {Graham} J.~R.,  eds, ~
  Vol. 299, Exploring the Formation and Evolution of Planetary Systems. pp
  271--272, \mn@doi{10.1017/S1743921313008545}

\bibitem[\protect\citeauthoryear{{Anders} et~al.,}{{Anders}
  et~al.}{2022}]{2022A&A...658A..91A}
{Anders} F.,  et~al., 2022, \mn@doi [\aap] {10.1051/0004-6361/202142369}, \href
  {https://ui.adsabs.harvard.edu/abs/2022A&A...658A..91A} {658, A91}

\bibitem[\protect\citeauthoryear{{Bailer-Jones}, {Rybizki}, {Fouesneau},
  {Demleitner}  \& {Andrae}}{{Bailer-Jones} et~al.}{2021}]{2021AJ....161..147B}
{Bailer-Jones} C.~A.~L.,  {Rybizki} J.,  {Fouesneau} M.,  {Demleitner} M.,
  {Andrae} R.,  2021, \mn@doi [\aj] {10.3847/1538-3881/abd806}, \href
  {https://ui.adsabs.harvard.edu/abs/2021AJ....161..147B} {161, 147}

\bibitem[\protect\citeauthoryear{{Bateson}}{{Bateson}}{1975}]{1975PVSS....3....1B}
{Bateson} F.~M.,  1975, Royal Astronomical Society of New Zealand Publications
  of Variable Star Section, \href
  {https://ui.adsabs.harvard.edu/abs/1975PVSS....3....1B} {3, 1}

\bibitem[\protect\citeauthoryear{{Bayo}, {Rodrigo}, {Barrado Y Navascu{\'e}s},
  {Solano}, {Guti{\'e}rrez}, {Morales-Calder{\'o}n}  \& {Allard}}{{Bayo}
  et~al.}{2008}]{2008A&A...492..277B}
{Bayo} A.,  {Rodrigo} C.,  {Barrado Y Navascu{\'e}s} D.,  {Solano} E.,
  {Guti{\'e}rrez} R.,  {Morales-Calder{\'o}n} M.,   {Allard} F.,  2008, \mn@doi
  [\aap] {10.1051/0004-6361:200810395}, \href
  {https://ui.adsabs.harvard.edu/abs/2008A&A...492..277B} {492, 277}

\bibitem[\protect\citeauthoryear{{Belczy{\'n}ski}, {Miko{\l}ajewska}, {Munari},
  {Ivison}  \& {Friedjung}}{{Belczy{\'n}ski}
  et~al.}{2000}]{2000A&AS..146..407B}
{Belczy{\'n}ski} K.,  {Miko{\l}ajewska} J.,  {Munari} U.,  {Ivison} R.~J.,
  {Friedjung} M.,  2000, \mn@doi [\aaps] {10.1051/aas:2000280}, \href
  {https://ui.adsabs.harvard.edu/abs/2000A&AS..146..407B} {146, 407}

\bibitem[\protect\citeauthoryear{{Buzzoni}, {Patelli}, {Bellazzini}, {Pecci}
  \& {Oliva}}{{Buzzoni} et~al.}{2010}]{2010MNRAS.403.1592B}
{Buzzoni} A.,  {Patelli} L.,  {Bellazzini} M.,  {Pecci} F.~F.,   {Oliva} E.,
  2010, \mn@doi [\mnras] {10.1111/j.1365-2966.2009.16223.x}, \href
  {https://ui.adsabs.harvard.edu/abs/2010MNRAS.403.1592B} {403, 1592}

\bibitem[\protect\citeauthoryear{{Cardelli}, {Clayton}  \& {Mathis}}{{Cardelli}
  et~al.}{1989}]{1989ApJ...345..245C}
{Cardelli} J.~A.,  {Clayton} G.~C.,   {Mathis} J.~S.,  1989, \mn@doi [\apj]
  {10.1086/167900}, \href
  {https://ui.adsabs.harvard.edu/abs/1989ApJ...345..245C} {345, 245}

\bibitem[\protect\citeauthoryear{{C{\'u}neo}, {Kenyon}, {G{\'o}mez}, {Chochol},
  {Shugarov}  \& {Kolotilov}}{{C{\'u}neo} et~al.}{2018}]{2018MNRAS.479.2728C}
{C{\'u}neo} V.~A.,  {Kenyon} S.~J.,  {G{\'o}mez} M.~N.,  {Chochol} D.,
  {Shugarov} S.~Y.,   {Kolotilov} E.~A.,  2018, \mn@doi [\mnras]
  {10.1093/mnras/sty1686}, \href
  {https://ui.adsabs.harvard.edu/abs/2018MNRAS.479.2728C} {479, 2728}

\bibitem[\protect\citeauthoryear{{Flower}}{{Flower}}{1996}]{1996ApJ...469..355F}
{Flower} P.~J.,  1996, \mn@doi [\apj] {10.1086/177785}, \href
  {https://ui.adsabs.harvard.edu/abs/1996ApJ...469..355F} {469, 355}

\bibitem[\protect\citeauthoryear{{Gaia Collaboration} et~al.,}{{Gaia
  Collaboration} et~al.}{2018}]{2018A&A...616A..10G}
{Gaia Collaboration} et~al., 2018, \mn@doi [\aap]
  {10.1051/0004-6361/201832843}, \href
  {https://ui.adsabs.harvard.edu/abs/2018A&A...616A..10G} {616, A10}

\bibitem[\protect\citeauthoryear{{Gaia Collaboration} et~al.,}{{Gaia
  Collaboration} et~al.}{2022}]{2022arXiv220800211G}
{Gaia Collaboration} et~al., 2022, \mn@doi [arXiv e-prints]
  {10.48550/arXiv.2208.00211}, \href
  {https://ui.adsabs.harvard.edu/abs/2022arXiv220800211G} {p. arXiv:2208.00211}

\bibitem[\protect\citeauthoryear{{Gromadzki}, {Miko{\l}ajewska}  \&
  {Soszy{\'n}ski}}{{Gromadzki} et~al.}{2013}]{2013AcA....63..405G}
{Gromadzki} M.,  {Miko{\l}ajewska} J.,   {Soszy{\'n}ski} I.,  2013, \mn@doi
  [\actaa] {10.48550/arXiv.1312.6063}, \href
  {https://ui.adsabs.harvard.edu/abs/2013AcA....63..405G} {63, 405}

\bibitem[\protect\citeauthoryear{{Henden}, {Levine}, {Terrell}  \&
  {Welch}}{{Henden} et~al.}{2015}]{2015AAS...22533616H}
{Henden} A.~A.,  {Levine} S.,  {Terrell} D.,   {Welch} D.~L.,  2015, in
  American Astronomical Society Meeting Abstracts \#225. p. 336.16

\bibitem[\protect\citeauthoryear{{I{\l}kiewicz}, {Miko{\l}ajewska},
  {Miszalski}, {Gromadzki}, {Monard}  \& {Amigo}}{{I{\l}kiewicz}
  et~al.}{2019}]{2019A&A...624A.133I}
{I{\l}kiewicz} K.,  {Miko{\l}ajewska} J.,  {Miszalski} B.,  {Gromadzki} M.,
  {Monard} B.,   {Amigo} P.,  2019, \mn@doi [\aap]
  {10.1051/0004-6361/201834165}, \href
  {https://ui.adsabs.harvard.edu/abs/2019A&A...624A.133I} {624, A133}

\bibitem[\protect\citeauthoryear{{Kato}, {Miko{\l}ajewska}  \&
  {Hachisu}}{{Kato} et~al.}{2012}]{2012ApJ...750....5K}
{Kato} M.,  {Miko{\l}ajewska} J.,   {Hachisu} I.,  2012, \mn@doi [\apj]
  {10.1088/0004-637X/750/1/5}, \href
  {https://ui.adsabs.harvard.edu/abs/2012ApJ...750....5K} {750, 5}

\bibitem[\protect\citeauthoryear{{Kilkenny}}{{Kilkenny}}{1989}]{1989Obs...109..229K}
{Kilkenny} D.,  1989, The Observatory, \href
  {https://ui.adsabs.harvard.edu/abs/1989Obs...109..229K} {109, 229}

\bibitem[\protect\citeauthoryear{Kloppenborg}{Kloppenborg}{2022}]{kafka}
Kloppenborg B.~K.,  2022, Observations from the AAVSO International Database,
  https://www.aavso.org

\bibitem[\protect\citeauthoryear{{Kochanek} et~al.,}{{Kochanek}
  et~al.}{2017}]{2017PASP..129j4502K}
{Kochanek} C.~S.,  et~al., 2017, \mn@doi [\pasp] {10.1088/1538-3873/aa80d9},
  \href {https://ui.adsabs.harvard.edu/abs/2017PASP..129j4502K} {129, 104502}

\bibitem[\protect\citeauthoryear{{Merc}, {G{\'a}lis}  \& {Wolf}}{{Merc}
  et~al.}{2019}]{2019RNAAS...3...28M}
{Merc} J.,  {G{\'a}lis} R.,   {Wolf} M.,  2019, \mn@doi [Research Notes of the
  American Astronomical Society] {10.3847/2515-5172/ab0429}, \href
  {https://ui.adsabs.harvard.edu/abs/2019RNAAS...3...28M} {3, 28}

\bibitem[\protect\citeauthoryear{{Merc}, {G{\'a}lis}, {K{\'a}ra}, {Wolf}  \&
  {Vra{\v{s}}{\v{t}}{\'a}k}}{{Merc} et~al.}{2020}]{2020MNRAS.499.2116M}
{Merc} J.,  {G{\'a}lis} R.,  {K{\'a}ra} J.,  {Wolf} M.,
  {Vra{\v{s}}{\v{t}}{\'a}k} M.,  2020, \mn@doi [\mnras]
  {10.1093/mnras/staa3063}, \href
  {https://ui.adsabs.harvard.edu/abs/2020MNRAS.499.2116M} {499, 2116}

\bibitem[\protect\citeauthoryear{{Merc} et~al.,}{{Merc}
  et~al.}{2021}]{2021MNRAS.506.4151M}
{Merc} J.,  et~al., 2021, \mn@doi [\mnras] {10.1093/mnras/stab2034}, \href
  {https://ui.adsabs.harvard.edu/abs/2021MNRAS.506.4151M} {506, 4151}

\bibitem[\protect\citeauthoryear{{Merc}, {G{\'a}lis}, {Wolf}, {Velez},
  {Bohlsen}  \& {Barlow}}{{Merc} et~al.}{2022}]{2022MNRAS.510.1404M}
{Merc} J.,  {G{\'a}lis} R.,  {Wolf} M.,  {Velez} P.,  {Bohlsen} T.,   {Barlow}
  B.~N.,  2022, \mn@doi [\mnras] {10.1093/mnras/stab3512}, \href
  {https://ui.adsabs.harvard.edu/abs/2022MNRAS.510.1404M} {510, 1404}

\bibitem[\protect\citeauthoryear{{Miko{\l}ajewska}}{{Miko{\l}ajewska}}{2003}]{2003ASPC..303....9M}
{Miko{\l}ajewska} J.,  2003, in {Corradi} R.~L.~M.,  {Mikolajewska} J.,
  {Mahoney} T.~J.,  eds,  Astronomical Society of the Pacific Conference Series
  Vol. 303, Symbiotic Stars Probing Stellar Evolution. p.~9 (\mn@eprint {arXiv}
  {astro-ph/0210489}), \mn@doi{10.48550/arXiv.astro-ph/0210489}

\bibitem[\protect\citeauthoryear{{Mikolajewska}}{{Mikolajewska}}{2010}]{2010arXiv1011.5657M}
{Mikolajewska} J.,  2010, \mn@doi [arXiv e-prints] {10.48550/arXiv.1011.5657},
  \href {https://ui.adsabs.harvard.edu/abs/2010arXiv1011.5657M} {p.
  arXiv:1011.5657}

\bibitem[\protect\citeauthoryear{{Miko{\l}ajewska}}{{Miko{\l}ajewska}}{2012}]{2012BaltA..21....5M}
{Miko{\l}ajewska} J.,  2012, \mn@doi [Baltic Astronomy]
  {10.1515/astro-2017-0352}, \href
  {https://ui.adsabs.harvard.edu/abs/2012BaltA..21....5M} {21, 5}

\bibitem[\protect\citeauthoryear{{Mikolajewska}, {Brandi}, {Hack}, {Whitelock},
  {Barba}, {Garcia}  \& {Marang}}{{Mikolajewska}
  et~al.}{1999}]{1999MNRAS.305..190M}
{Mikolajewska} J.,  {Brandi} E.,  {Hack} W.,  {Whitelock} P.~A.,  {Barba} R.,
  {Garcia} L.,   {Marang} F.,  1999, \mn@doi [\mnras]
  {10.1046/j.1365-8711.1999.02449.x}, \href
  {https://ui.adsabs.harvard.edu/abs/1999MNRAS.305..190M} {305, 190}

\bibitem[\protect\citeauthoryear{{Munari}}{{Munari}}{2019}]{2019arXiv190901389M}
{Munari} U.,  2019, \mn@doi [arXiv e-prints] {10.48550/arXiv.1909.01389}, \href
  {https://ui.adsabs.harvard.edu/abs/2019arXiv190901389M} {p. arXiv:1909.01389}

\bibitem[\protect\citeauthoryear{{Murset} \& {Nussbaumer}}{{Murset} \&
  {Nussbaumer}}{1994}]{1994A&A...282..586M}
{Murset} U.,  {Nussbaumer} H.,  1994, \aap, \href
  {https://ui.adsabs.harvard.edu/abs/1994A&A...282..586M} {282, 586}

\bibitem[\protect\citeauthoryear{{M{\"u}rset} \& {Schmid}}{{M{\"u}rset} \&
  {Schmid}}{1999}]{1999A&AS..137..473M}
{M{\"u}rset} U.,  {Schmid} H.~M.,  1999, \mn@doi [\aaps] {10.1051/aas:1999105},
  \href {https://ui.adsabs.harvard.edu/abs/1999A&AS..137..473M} {137, 473}

\bibitem[\protect\citeauthoryear{{Nomoto}, {Saio}, {Kato}  \&
  {Hachisu}}{{Nomoto} et~al.}{2007}]{2007ApJ...663.1269N}
{Nomoto} K.,  {Saio} H.,  {Kato} M.,   {Hachisu} I.,  2007, \mn@doi [\apj]
  {10.1086/518465}, \href
  {https://ui.adsabs.harvard.edu/abs/2007ApJ...663.1269N} {663, 1269}

\bibitem[\protect\citeauthoryear{{Pickles}}{{Pickles}}{1998}]{1998PASP..110..863P}
{Pickles} A.~J.,  1998, \mn@doi [\pasp] {10.1086/316197}, \href
  {https://ui.adsabs.harvard.edu/abs/1998PASP..110..863P} {110, 863}

\bibitem[\protect\citeauthoryear{{Pojmanski}}{{Pojmanski}}{1997}]{1997AcA....47..467P}
{Pojmanski} G.,  1997, \actaa, \href
  {https://ui.adsabs.harvard.edu/abs/1997AcA....47..467P} {47, 467}

\bibitem[\protect\citeauthoryear{{Samus'}, {Kazarovets}, {Durlevich}, {Kireeva}
   \& {Pastukhova}}{{Samus'} et~al.}{2017}]{2017ARep...61...80S}
{Samus'} N.~N.,  {Kazarovets} E.~V.,  {Durlevich} O.~V.,  {Kireeva} N.~N.,
  {Pastukhova} E.~N.,  2017, \mn@doi [Astronomy Reports]
  {10.1134/S1063772917010085}, \href
  {https://ui.adsabs.harvard.edu/abs/2017ARep...61...80S} {61, 80}

\bibitem[\protect\citeauthoryear{{Schlafly} \& {Finkbeiner}}{{Schlafly} \&
  {Finkbeiner}}{2011}]{2011ApJ...737..103S}
{Schlafly} E.~F.,  {Finkbeiner} D.~P.,  2011, \mn@doi [\apj]
  {10.1088/0004-637X/737/2/103}, \href
  {https://ui.adsabs.harvard.edu/abs/2011ApJ...737..103S} {737, 103}

\bibitem[\protect\citeauthoryear{{Shappee} et~al.,}{{Shappee}
  et~al.}{2014}]{2014ApJ...788...48S}
{Shappee} B.~J.,  et~al., 2014, \mn@doi [\apj] {10.1088/0004-637X/788/1/48},
  \href {https://ui.adsabs.harvard.edu/abs/2014ApJ...788...48S} {788, 48}

\bibitem[\protect\citeauthoryear{{Skopal}}{{Skopal}}{2005}]{2005A&A...440..995S}
{Skopal} A.,  2005, \mn@doi [\aap] {10.1051/0004-6361:20034262}, \href
  {https://ui.adsabs.harvard.edu/abs/2005A&A...440..995S} {440, 995}

\bibitem[\protect\citeauthoryear{{Skopal} et~al.,}{{Skopal}
  et~al.}{2020}]{2020A&A...636A..77S}
{Skopal} A.,  et~al., 2020, \mn@doi [\aap] {10.1051/0004-6361/201937199}, \href
  {https://ui.adsabs.harvard.edu/abs/2020A&A...636A..77S} {636, A77}

\bibitem[\protect\citeauthoryear{{Skrutskie} et~al.,}{{Skrutskie}
  et~al.}{2006}]{2006AJ....131.1163S}
{Skrutskie} M.~F.,  et~al., 2006, \mn@doi [\aj] {10.1086/498708}, \href
  {https://ui.adsabs.harvard.edu/abs/2006AJ....131.1163S} {131, 1163}

\bibitem[\protect\citeauthoryear{{Swope} \& {Shapley}}{{Swope} \&
  {Shapley}}{1940}]{1940AnHar..90..207S}
{Swope} H.~H.,  {Shapley} H.,  1940, Annals of Harvard College Observatory,
  \href {https://ui.adsabs.harvard.edu/abs/1940AnHar..90..207S} {90, 207}

\bibitem[\protect\citeauthoryear{{Teyssier}}{{Teyssier}}{2019}]{2019CoSka..49..217T}
{Teyssier} F.,  2019, Contributions of the Astronomical Observatory Skalnate
  Pleso, \href {https://ui.adsabs.harvard.edu/abs/2019CoSka..49..217T} {49,
  217}

\bibitem[\protect\citeauthoryear{{Thackeray}}{{Thackeray}}{1977}]{1977MmRAS..83....1T}
{Thackeray} A.~D.,  1977, \memras, \href
  {https://ui.adsabs.harvard.edu/abs/1977MmRAS..83....1T} {83, 1}

\bibitem[\protect\citeauthoryear{{Wenger} et~al.,}{{Wenger}
  et~al.}{2000}]{2000A&AS..143....9W}
{Wenger} M.,  et~al., 2000, \mn@doi [\aaps] {10.1051/aas:2000332}, \href
  {https://ui.adsabs.harvard.edu/abs/2000A&AS..143....9W} {143, 9}

\bibitem[\protect\citeauthoryear{{Wolf}, {Bildsten}, {Brooks}  \&
  {Paxton}}{{Wolf} et~al.}{2013}]{2013ApJ...777..136W}
{Wolf} W.~M.,  {Bildsten} L.,  {Brooks} J.,   {Paxton} B.,  2013, \mn@doi
  [\apj] {10.1088/0004-637X/777/2/136}, \href
  {https://ui.adsabs.harvard.edu/abs/2013ApJ...777..136W} {777, 136}

\bibitem[\protect\citeauthoryear{{Wright} et~al.,}{{Wright}
  et~al.}{2010}]{2010AJ....140.1868W}
{Wright} E.~L.,  et~al., 2010, \mn@doi [\aj] {10.1088/0004-6256/140/6/1868},
  \href {https://ui.adsabs.harvard.edu/abs/2010AJ....140.1868W} {140, 1868}

\bibitem[\protect\citeauthoryear{{de Jager} \& {Nieuwenhuijzen}}{{de Jager} \&
  {Nieuwenhuijzen}}{1987}]{1987A&A...177..217D}
{de Jager} C.,  {Nieuwenhuijzen} H.,  1987, \aap, \href
  {https://ui.adsabs.harvard.edu/abs/1987A&A...177..217D} {177, 217}

\bibitem[\protect\citeauthoryear{{van Belle} et~al.,}{{van Belle}
  et~al.}{1999}]{1999AJ....117..521V}
{van Belle} G.~T.,  et~al., 1999, \mn@doi [\aj] {10.1086/300677}, \href
  {https://ui.adsabs.harvard.edu/abs/1999AJ....117..521V} {117, 521}

\makeatother
\end{thebibliography}




\appendix

\begin{table}\section{Log of observations}
\caption{Log of observations. The star symbol (*) denotes the spectra shown in Fig. \ref{fig:spectra}. Observer codes: VLZ, CUR - P. Velez, S. Curry (32-cm Planewave CDK telescope, UVEX spectrograph; Australia); 2SPOT - S. Charbonnel, O. Garde, P. Le~D\^u, L. Mulato, T. Petit (30-cm Ritchey-Chretien telescope, Alpy600 spectrograph; Chile); BHQ - T. Bohlsen (28-cm Celestron telescope, LISA spectrograph; Australia); TLO - T. Love (30-cm Ritchey-Chretien telescope, Alpy600 spectrograph; New Zealand); HBA - H. Barker (20-cm Newtonian telescope, L200 Littrow spectrograph; New Zealand).}             
\label{table:log_obs}      
\centering
\begin{tabular}{lllcc}
\hline\hline
JD& Resolution & $\lambda_{\rm min}$-$\lambda_{\rm max}$ & Observer \\
2\,459\,.. &  & [\AA] &  \\\hline
*061.01 & 1906 & 3800-8000 & BHQ \\
523.92 & 1156 & 4754-6788 & VLZ \\
*678.80 & 566 & 3750-7807 & 2SPOT \\
680.23 & 1344 & 4658-6835 & VLZ \\
701.21 & 1402 & 4690-6751 & VLZ \\
705.19 & 1445 & 4705-6764 & VLZ \\
708.16 & 1420 & 4697-6757 & VLZ \\
716.17 & 1237 & 3659-5584 & VLZ \\
717.14 & 878 & 3671-5580 & VLZ \\
*719.71 & 571 & 3603-7809 & 2SPOT \\
732.27 & 851 & 3674-5585 & VLZ \\
753.17 & 867 & 3650-5583 & VLZ \\
775.12 & 761 & 3700-5500 & CUR \\
786.01 & 1503 & 3800-7000 & BHQ \\
787.54 & 589 & 3700-7745 & 2SPOT \\
790.94 & 1526 & 3800-7000 & BHQ \\
*795.62 & 568 & 3603-7808 & 2SPOT \\
796.93 & 614 & 3801-7501 & TLO \\
799.05 & 1116 & 4661-6735 & CUR \\
807.71 & 433 & 3701-7750 & 2SPOT \\
809.00 & 1119 & 4656-6690 & CUR \\
815.94 & 612 & 3801-7501 & TLO \\
822.99 & 1147 & 4669-6705 & VLZ \\
*824.51 & 570 & 3603-7404 & 2SPOT \\
874.92 & 867 & 3756-7359 & HBA \\
875.89 & 607 & 3801-7501 & TLO \\
879.93 & 1101 & 3639-5545 & VLZ \\
880.92 & 1266 & 4652-6688 & VLZ \\
881.92 & 1610 & 6493-8549 & VLZ \\
892.95 & 3216 & 4501-5150 & VLZ \\
986.26 & 1195 & 4708-6746 & VLZ \\
987.26 & 1038 & 3749-5609 & VLZ \\
988.26 & 1708 & 6548-8590 & VLZ \\
990.25 & 1712 & 6555-8600 & VLZ \\
992.25 & 1672 & 6550-8591 & VLZ\\\hline
\end{tabular}
\end{table}


\bsp	
\label{lastpage}
\end{document}